\documentclass[sigconf]{acmart}

\usepackage{censor}
\usepackage{graphicx}
\usepackage{subfig}
\usepackage{amsmath}
\usepackage{algorithm}
\usepackage{algpseudocode}
\usepackage[utf8]{inputenc}

\def\BibTeX{{\rm B\kern-.05em{\sc i\kern-.025em b}\kern-.08emT\kern-.1667em\lower.7ex\hbox{E}\kern-.125emX}}
    
\copyrightyear{2018}
\acmYear{2018}
\setcopyright{none}
\acmConference[RecSys '19]{RecSys '19: The ACM Conference Series on Recommender Systems}{September 16--20, 2019}{Copenhagen}
\acmBooktitle{RecSys '19: The ACM Conference Series on Recommender Systems, September 16--20, 2019, Copenhagen, Denmark}
\acmPrice{15.00}
\acmDOI{10.1145/1122445.1122456}
\acmISBN{978-1-4503-9999-9/18/06}

\renewcommand\footnotetextcopyrightpermission[1]{} 
\begin{document}

%
\title{SeER: An Explainable Deep Learning MIDI-based Hybrid Song Recommender System}

%

\author{Khalil Damak}
\affiliation{%
  \institution{Knowledge Discovery and Web Mining Lab, CECS Department, University of Louisville}
  \streetaddress{Louisville, KY 40292}
  \city{Louisville}
  \country{USA}}
\email{khalil.damak@louisville.edu}

\author{Olfa Nasraoui}
\affiliation{%
  \institution{Knowledge Discovery and Web Mining Lab, CECS Department, University of Louisville}
  \streetaddress{Louisville, KY 40292}
  \city{Louisville}
  \country{USA}}
\email{olfa.nasraoui@louisville.edu}

%

\renewcommand{\shortauthors}{Damak and Nasraoui}

%
\begin{abstract}
State of the art music recommender systems mainly rely on either matrix factorization-based collaborative filtering approaches or deep learning architectures. Deep learning models usually use metadata for content-based filtering or predict the next user interaction by learning from temporal sequences of user actions. Despite advances in deep learning for song recommendation, none has taken advantage of the sequential nature of songs by learning sequence models that are based on content.
Aside from the importance of prediction accuracy, other significant aspects are important, such as explainability and solving the cold start problem.
In this work, we propose a hybrid deep learning model, called ``SeER", that uses collaborative filtering (CF) and deep learning sequence models on the MIDI content of songs for recommendation in order to provide more accurate personalized recommendations; solve the item cold start problem; and generate a relevant explanation for a song recommendation.
Our evaluation experiments show promising results compared to state of the art baseline and hybrid song recommender systems in terms of ranking evaluation. Moreover, based on proposed tests for offline validation, we show that our personalized explanations capture properties that are in accordance with the user's preferences.
\end{abstract}

%
%

\settopmatter{printacmref=false}

%

%
%
\maketitle

\section{Introduction}

Recommendation is becoming a prevalent component of our daily lives that has attracted increasing interest from the Machine Learning research community in recent years. Among the fields in which recommendation is most decisive is music. Music streaming platforms are indeed numerous: Spotify \cite{spotify}, Pandora \cite{pandora}, YouTube Music \cite{youtube_music} and many others. However, what makes the success of a platform is its capacity to predict which song the user wants to listen to at the moment given their previous interactions. The most accurate recommender systems rely on complex black box machine learning models that do not explain why they output the predicted recommendation. In fact, one main challenge is designing a recommender system that mitigates the trade-off between explainability and prediction accuracy ~\cite{Abdollahi:2017:UEC:3109859.3109913}. Today, the most widely used techniques in music recommendation are matrix factorization (MF)-based collaborative filtering approaches \cite{8066567} and deep learning architectures \cite{DBLP:journals/corr/ZhangYS17aa}. MF is based on similarities between users and items in a latent space obtained by factorizing the rating matrix into user and item latent factor matrices ~\cite{Koren:2009:MFT:1608565.1608614}. For state of the art deep learning recommender systems, there are mainly two approaches. The first approach relies on content based filtering \cite{Meteren2000UsingCF} using metadata to recommend items. The second approach uses sequence models \cite{Lipton2015ACR} \cite{D14-1179} \cite{Hochreiter:1997:LSM:1246443.1246450} to predict the next interaction (played song) given the previous interactions \cite{hidasi2015session}\cite{DBLP:journals/corr/TanXL16}\cite{Wu2016PersonalRU}. Despite the advances in deep learning for song recommendation and despite the sequential nature of songs that makes them naturally adapted to sequence models, no work has used sequence models with the \textbf{content} of songs for recommendation. Aside from accuracy and explainability, the cold start problem is a significant issue for collaborative filtering recommender systems \cite{abdollahi_2017}. In fact, most recommender systems need an initial history of interactions (ratings, clicks, plays, etc.) to recommend items. In music streaming platforms, new users and songs are constantly added making solving this issue crucial.
In this work, we take advantage of the sequential nature of the songs, the prediction power of MF and the superior capabilities of deep learning sequence models to achieve the following objectives:
\begin{itemize}
    \item Propose a method to transform the Musical Instrument Digital Interface (MIDI) format \cite{web_midi} of songs into multidimentional time series to be used as input to deep learning sequence models and keep a large amount of information about the song;
    \item Integrate content based filtering using deep learning sequence models into collaborative filtering MF to build a novel hybrid model that provides accurate predictions compared to baseline recommender systems, solves the item cold start problem, and provides explanations to the recommendations; and
    \item Propose a new type of explanation to song recommendation that consists of presenting to the user a short personalized MIDI segment of the song that characterizes the portion that the user is predicted to like the most.
\end{itemize}

\section{Related work}

Various recommender systems rely on sequence models. However, not all of them use them for recommendation with user preferences. In fact, some are session-based CF models \cite{hidasi2015session}\cite{DBLP:journals/corr/TanXL16}\cite{Wu2016PersonalRU} that predict the next interaction in a sequence of interactions regardless of the user's personal preferences. Other methods introduce content to session-based recommendation \cite{Hidasi:2016:PRN:2959100.2959167}\cite{DBLP:journals/corr/SmirnovaV17} and prove that side information enhances the recommendation quality \cite{DBLP:journals/corr/ZhangYS17aa}. Other recommender systems using sequence models take into consideration user identification \cite{Wu:2017:RRN:3018661.3018689}\cite{joint_training}. These engines model temporal dependencies for both users and movies \cite{Wu:2017:RRN:3018661.3018689}\cite{joint_training} and generate reviews \cite{joint_training}. The main objective of these models is to predict ratings of users to items using seasonal evolutions of items and user preferences in addition to user and item latent vectors. Alternate models aimed to generate review tips \cite{DBLP:journals/corr/abs-1708-00154}, predict the returning time of users and predict items \cite{Jing:2017:NSR:3018661.3018719} or produce next item recommendations for a user by proposing a novel Gated Recurrent Unit \cite{D14-1179} (GRU) structure \cite{Donkers:2017:SUR:3109859.3109877}. Finally, some recommender systems also use sequence models as a feature representation learning tool \cite{DBLP:journals/corr/ZhangYS17aa}. \cite{bansal2016multitask} creates a latent representation of items and uses it as input to a CF model with a user embedding to predict ratings. 
On the other hand, song recommendation received contributions from few hybrid models that often diverge in terms of input data and features created. In fact, music items can be represented by features derived from audio signals, social tags or web content \cite{DBLP:journals/corr/abs-1807-05858}. 
Among the most noticeable hybrid song recommender systems, \cite{Wang:2014:ICH:2647868.2654940} learns latent factors of users and items using matrix factorization and sums their product with the product obtained with created user and song features. \cite{Benzi2016SongRW} combines non-negative MF and graph regularization to predict the inclusion of a song in a playlist. \cite{DBLP:journals/corr/OramasNSS17} learns artist embeddings from biographies and track embeddings from audio spectrograms, then aggregates and multiplies them by user latent factors obtained by weighted MF to predict ratings. \cite{van2013deep} trains a Convolutional Neural Network \cite{lecun1999object} on spectrograms of song samples to predict latent features obtained with an MF approach for songs with no ratings. Finally, \cite{andjelkovic2018} positions the users in a mood space, given their favorite artists, and recommends new artists using similarity measures.

\section{Methods}

In this section, we start by describing the data that we used along with its preparation procedure. Then, we present our model called ``Sequence-based Explainable Recommender system" (SeER). This being done, we describe our explainability process called ``Segment Forward propagation". But first, we show the list of variables that are used in this section in Table~\ref{tab:notation_sec_4} to ease the reading of the remainder of this article.

\begin{table}
\caption{Notation used in Section 3. \label{tab:notation_sec_4} }
\begin{center}
\resizebox{\linewidth}{!}{
\begin{tabular}{ l c c c c c}
\hline
Symbol/Notation & Definition\\ \hline \hline
$U$ & User embedding (latent) matrix\\
$U_{u}$ & User $u$'s latent vector\\
$S$ & Song lookup matrix\\
$S_{s}$ & Song $s$'s flattened array\\
$x^{s}$ & Song $s$'s array (multidimensional time series)\\
$x^{s}_{t}$ & Song $s$'s array at time step $t$\\
$x^{s, k}$ & Array of segment $k$ from song $s$\\
$x^{s, exp}_{u}$ & Explainability segment array of song recommendation $s$ to user $u$\\
$R$ & Training data\\
$r_{us}$ & Actual rating of user $u$ to song $s$\\
$\hat{r}_{us}$ & Predicted rating of user $u$ to song $s$\\
$\hat{r}_{us}^{k}$ & Predicted rating of user $u$ to segment $k$ of song $s$\\
$h^{<m>, s}_{t}$ & Hidden state of sequence model layer $m$ at time step $t$ on input song $s$\\
$T$ & Normalized number of time steps\\
$|.|$ & Cardinality\\
$\;\Dot{}$ & Dot product\\
$\nabla_{w}J$ & Gradient of $J$ with respect to $w$\\
\hline
\end{tabular}
}
\end{center}
\end{table}

\subsection{Data Preparation}

\begin{figure}
\centering
\includegraphics[height=3.5cm]{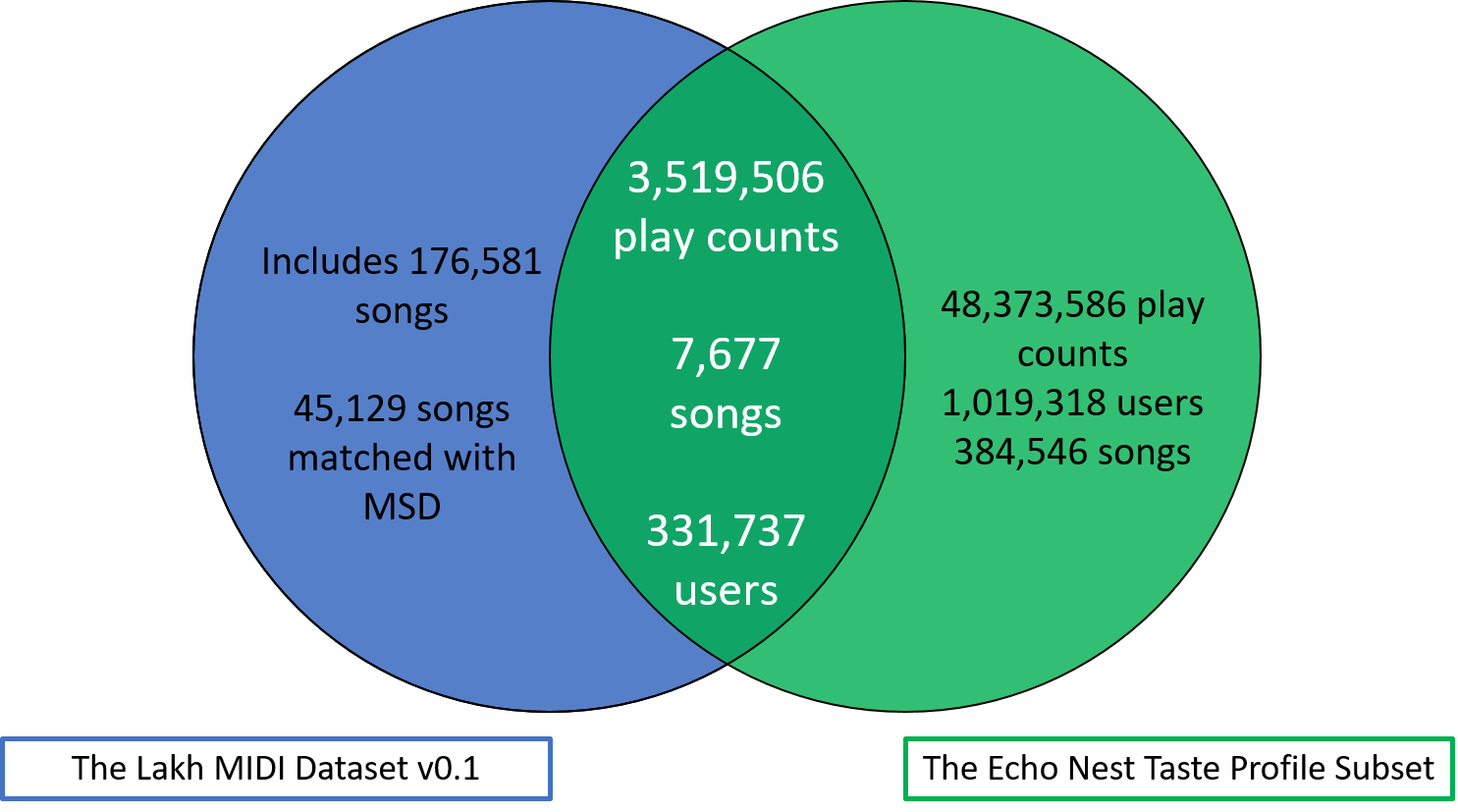}
\caption{ Our dataset resulting from the intersection between ``The Lakh MIDI Dataset v0.1" and ``The Echo Nest Taste Profile Subset". \label{fig:dataset} }
\end{figure}

We needed a dataset that includes both user to item interactions and song content data. Thus, we used two datasets from the Million Song Dataset (MSD) \cite{Bertin-Mahieux2011}. The Echo Nest Taste Profile Subset \cite{data_echonest} includes 48,373,586 play counts of 1,019,318 users to 384,546 songs collected from The Echo Nest's undisclosed partners. The Lakh MIDI Dataset includes 45,129 unique MIDI files matched to MSD songs \cite{data_lakh} \cite{raffel_2016}. We combined both datasets by taking the intersection in terms of songs as presented in Fig.~\ref{fig:dataset}. Then, we followed the same methodology used in \cite{He:2017:NCF:3038912.3052569} to reduce the sparsity of the data. We filtered out users that interacted with less than 20 unique songs. We obtained a dataset with 32,180 users, 6,442 songs with available MIDI files, and 941,044 play counts. Our dataset has a sparsity of 99.54\%.

\begin{figure}[!tbp]
  \centering
  \subfloat[]{\includegraphics[width=0.135\textwidth]{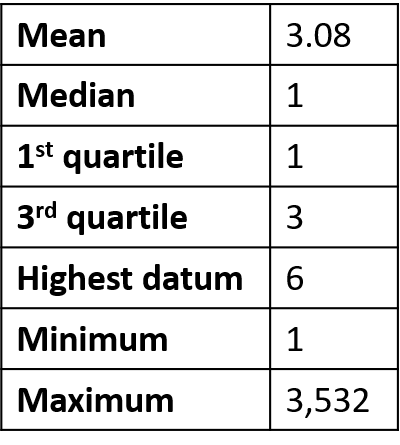}\label{fig:f1}}
  \hfill
  \subfloat[]{\includegraphics[width=0.25\textwidth]{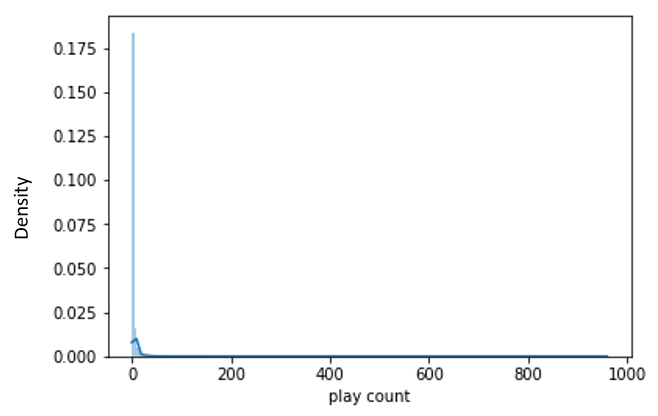}\label{fig:f2}}
  \caption{Play count statistics: (a) represents the statistics of the play count and (b) represents the density plot of the play count (filtered play counts < 1000 for better visualization).}
  \label{fig:play_count_visualization}
\end{figure}

\begin{figure}
\centering
\includegraphics[height=2cm]{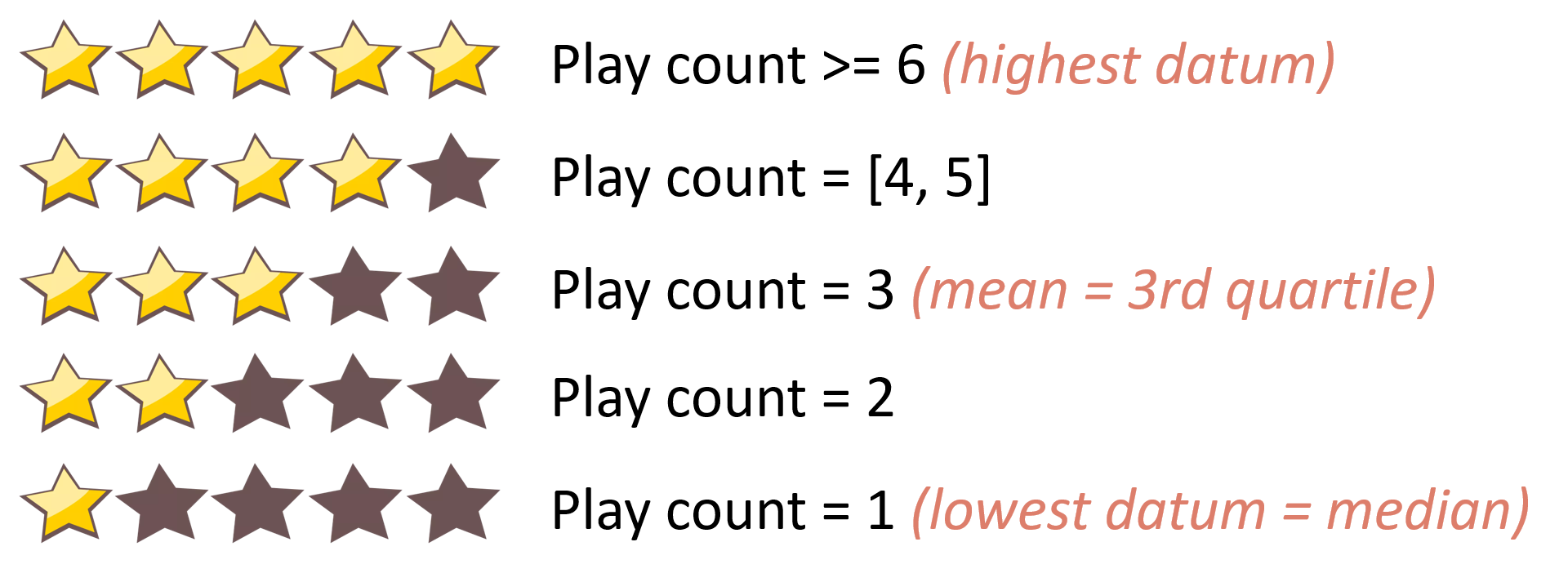}
\caption{ Play count normalization into 5-star ratings. \label{fig:play_count_normalization} }
\end{figure}

\begin{figure}
\centering
\includegraphics[height=3cm]{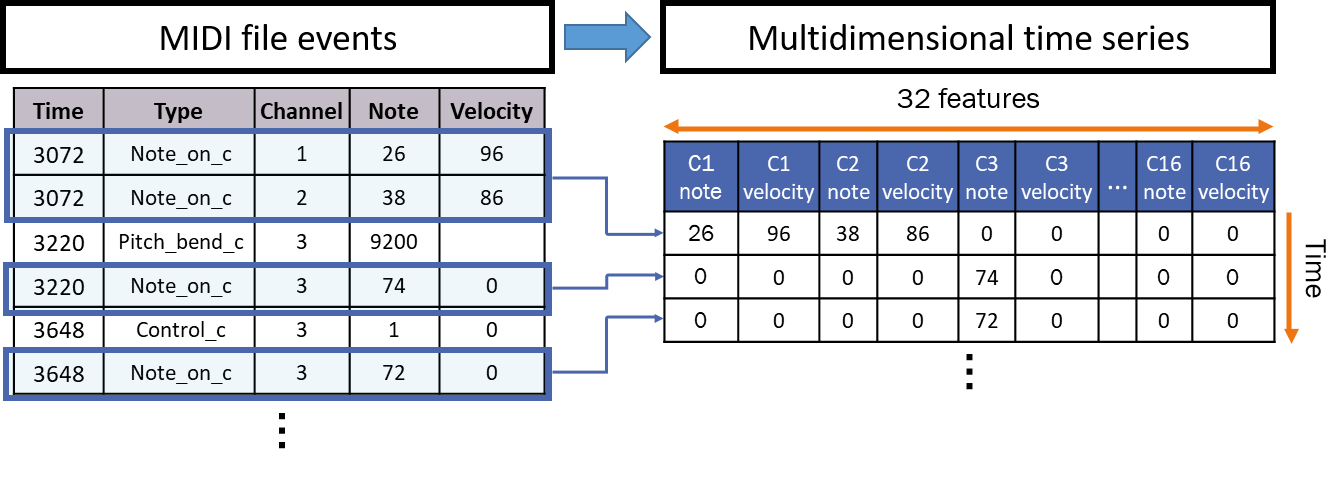}
\caption{ MIDI to multidimensional time series transformation process. \label{fig:midi_to_time_series} }
\end{figure}

We preprocessed our dataset by first mapping the play counts to ratings in order to remove outliers. To prove the necessity of this step, we show the distribution and statistics of the play counts in Fig.~\ref{fig:play_count_visualization}.
The play counts follow a power law distribution with a median of 1. Also, there are users that listened to the same song hundreds and thousands of times; and the maximum play count is 3,532. These high play counts are outliers that may bias the training of the model. While it is true that the more a song is listened to by a user, the more likely the user likes it, whether a user listens to a song 10 or 3,000 times, it is clear that they like it. Hence, both cases should be considered the same. Therefore, we used the statistics of the play counts to map them to ratings as shown in Fig.~\ref{fig:play_count_normalization}. Next, we created the input to train sequence models by transforming each MIDI file into a multidimensional time series. MIDI files are polyphonic digital instrumental audios that are usually used to create music. They are constituted of event messages that are consecutive in time \cite{web_midi}. Each message includes a type (such as a note), notation (the note played), time (the time it is played) and velocity (how rapidly and forcefully it is played) \cite{midicsv}. These events are distributed over 16 available channels of information, which are independent paths over which messages travel \cite{web_midi}. Each channel can be programmed to play one instrument. Thus, a MIDI file can play up to 16 instruments simultaneously. We first used ``MIDICSV" \cite{midicsv} to translate the MIDI files into sheets of the event messages.
We only considered the ``Note on C" events to focus our interest on the sequences of notes played throughout time. Thus, we extracted the notes that are played within the 16 channels with their velocities. As a result, each transformed multidimensional time series is constituted of a certain number of rows representing the number of ``Note on C" events and 32 features representing the notes and velocities played within the 16 channels. The transformation process is summarized in Fig.~\ref{fig:midi_to_time_series}. We then normalized the number of time steps to the median number of timesteps of the songs in our dataset (2,600) to be able to train with mini-batches \cite{Li:2014:EMT:2623330.2623612}. At least 50\% of the songs kept all their notes and 75\% of the songs kept at least half of their notes. Finally, in order to avoid duplicates of the same song in the input and ensure memory efficiency, we created a song lookup matrix by flattening each multidimensional time series into one row in the matrix.

\subsection{SeER}

\begin{figure}
\centering
\includegraphics[width=\linewidth]{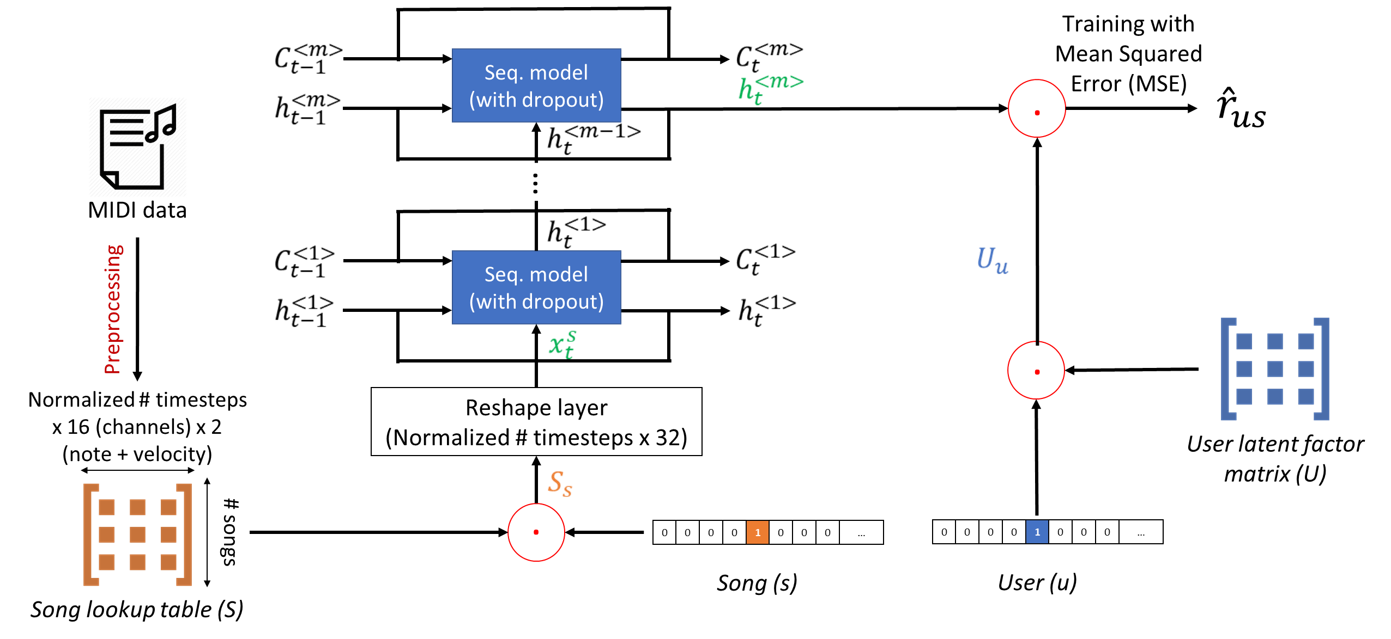}
\caption{ Structure of SeER: For every training tuple (user, song, rating), the model extracts the corresponding user latent vector and flat song array from the user latent matrix and the song lookup matrix respectively. The song array is reshaped to its original 2-dimensional format and input to a sequence model. The resulting song hidden state vector is multiplied with the user latent vector to predict the rating. \label{fig:model} }
\end{figure}

Three main observations motivated the design of our model. \textbf{First}, the sequential nature of songs, particularly represented by MIDI files, can be best modeled using sequential models. \textbf{Second}, the hidden state (output) of a sequence model is both learnable and of chosen size, both being basic properties of an embedding matrix. Thus, we opted to assimilate it to a user embedding. \textbf{Third}, sequence models can propagate instances with varying time steps. This inspired us to try to explain the recommendations using song segments. These motivations led us to design our model, called ``SeER": a sequence-based explainable hybrid song recommender system with the structure presented in Fig.~\ref{fig:model}. SeER takes as input the song lookup matrix and a user embedding matrix. For each user, song, and rating triplet $(u, s, r_{us})$ in the training data $R$, we extract the corresponding latent factor vector $U_{u}$ of the user and the flattened song array $S_{s}$. The latter process is illustrated in Fig.~\ref{fig:model} with multiplications of the user embedding and song lookup matrices with one hot vectors of $u$ and $s$ respectively. The song array is next reshaped into its two-dimensional original shape (2600 time steps by 32 features). The resulting array $x_{s}$ is input to a sequence model and, finally, the hidden state of the last layer $m$ at the last time step $(T = 2,600)$ $h^{<m>, s}_{T}$ is multiplied with the song latent vector $U_{u}$ to predict a rating of the user to the song $\hat{r}_{us} = U_{u} \cdot h^{<m>, s}_{T}$. To be consistent, we chose the size of the sequential hidden state to be the same as the number of user latent features. This enables computing the scalar product of the two latent vectors to yield a predicted rating. The model is trained using the Mean Squared Error (MSE) \cite{lehmann1998} as a loss function by comparing the actual rating $r_{us}$ to the predicted rating $\hat{r}_{us}$. Thus, our objective function is:

\begin{equation}
    J = \frac{1}{|R|} \sum_{(u, s, r_{us}) \epsilon \mathrm{R}}(\hat{r}_{us} - r_{us})^{2} = \frac{1}{|R|} \sum_{(u, s, r_{us}) \epsilon \mathrm{R}}(U_{u} \cdot h_{T}^{<m>, s} - r_{us})^{2}
\end{equation}

Note that in Fig.~\ref{fig:model}, the cell states can be ignored when using Recurrent Neural Networks \cite{Lipton2015ACR} (RNNs) or GRUs.

The training process of SeER for each epoch is described in Alg.~\ref{alg:SeER}.

\begin{algorithm}
  \caption{SeER training algorithm (for each epoch) with step size $\alpha$, using mini-batch gradient descent for simplicity}\label{alg:SeER}
  \begin{algorithmic}[1]
    \Procedure{SeER\_training\_epoch }{song lookup matrix $S$, user latent factor matrix $U$, set of mini-batches $B$, learning rate $\alpha$, number of sequence model layers $m$, number of timesteps $T$}
      \For{$b\:in\:B$}
        \For{$(u, s, r_{us})\:in\:b$}
            \State $U_{u} \gets One\_hot(u) \cdot U$\Comment{extract latent vector of $u$}
            \State $S_{s} \gets One\_hot(s) \cdot S$\Comment{extract flat song array of $s$}
            \State $x^{s} \gets Reshape(S_{s})$\Comment{reshape $S_{s}$ to $(T \times 32)$}
            \State $h_{T}^{<m>, s} \gets Sequence\_model(x^{s})$
            \State $\hat{r}_{us} \gets U_{u} \cdot h_{T}^{<m>, s}$\Comment{predict rating of $u$ to $s$}
        \EndFor
        \State $J=\frac{1}{|b|} \sum_{(u, s, r_{us}) \epsilon b}(U_{u} h_{T}^{<m>, s} - r_{us})^{2}$\Comment{compute prediction loss}
        \State $w \gets w - \alpha \cdot \nabla_{w} J$ \Comment{back propagate ($w$ refers to the parameters of $U$ and the sequence model)}
      \EndFor
    \EndProcedure
  \end{algorithmic}
\end{algorithm}

\subsection{Recommendation and Segment Forward Propagation Explainabiliy}

\begin{figure}
\centering
\includegraphics[height=2.75cm]{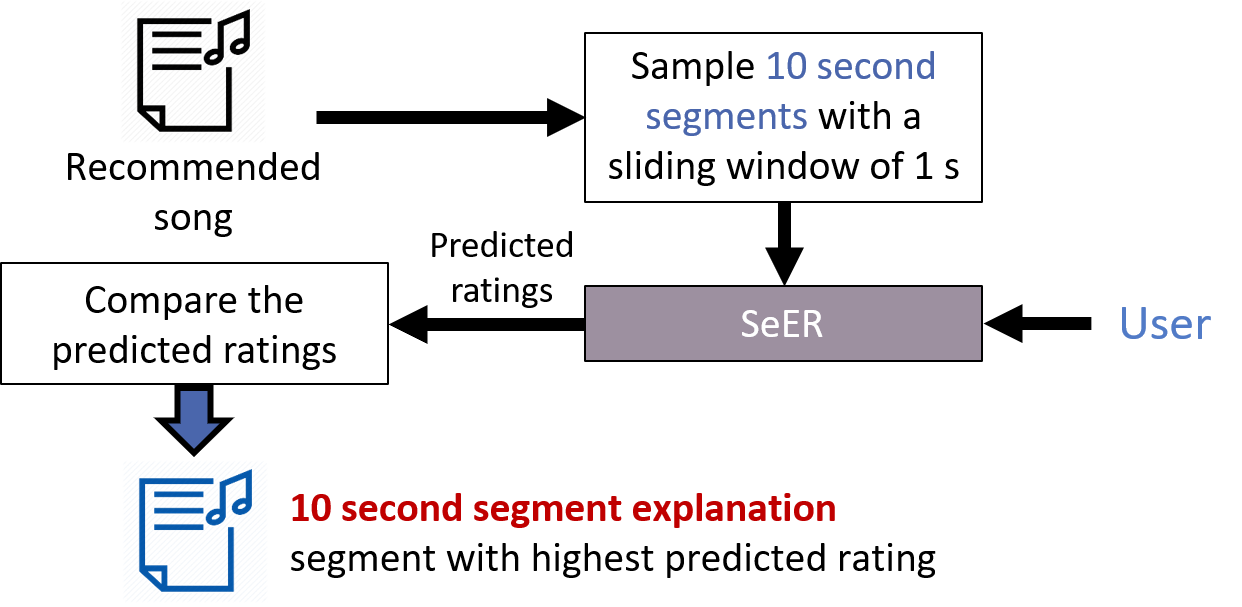}
\caption{ Segment Forward Propagation Explainability. \label{fig:explainability_process} }
\end{figure}

The recommendation consists of feeding user and unrated song inputs to the SeER model in Fig.~\ref{fig:model} which results in a predicted rating for each input song. The highest predicted ratings yield a list of recommended songs.
After generating a song recommendation $s$ to a user $u$, we explain it by presenting a 10-second MIDI segment $x^{s, exp}_{u}$ of the song that tries to capture the most important portion of the recommended song for the input user. First, we sample segments of the MIDI file by using a 10-second sliding window of one-second stride. This means that the first segment is the first 10 seconds of the audio, the second segment is from second 2 to second 11, and so forth, until we reach the end of the song. To do this, we start by creating absolute time segments that we match to MIDI times in the song to determine the range of time steps of each segment. In fact, the time in a MIDI file is in pulses and can be converted to absolute time such that $time[\mu s] =  \frac{MIDI\,time[pulses]}{Division[pulses/QR.\,note]} Tempo[\mu s/QR.\,note]$. The division is the number of pulses per quarter note and the tempo is a measure of speed \cite{midicsv}. Then, we create a multidimensional time series $x^{s, k}$ for each segment $k$ by truncating the time series of the recommended song $x^{s}$. Finally, we feed each segment's time series $x^{s, k}$ along with the user latent vector $U_{u}$ as input to the SeER model to estimate a rating $\hat{r}_{us}^{k}$ of that user to the segment. The segment that obtains the highest predicted rating $\hat{r}_{us}^{k}$ is presented to the user as an explanation for the song recommendation. We call this explainability process ``Segment Forward Propagation Explainability" because it relies on forward propagation of segments to explain the prediction. The aforementioned explanation process is presented in Fig.~\ref{fig:explainability_process} and is summarized in Alg.~\ref{alg:explainability}.

\begin{algorithm}
  \caption{Segment Forward Propagation Explainability}\label{alg:explainability}
  \begin{algorithmic}[1]
    \Procedure{Segment\_Forward\_Propagation }{recommended song $s$, length of $s$ in seconds $L$, song array $x^{s}$, user latent vector $U_{u}$, number of timesteps $T$, trained model $SeER$}
    \State $abs\_time\_x^{s} \gets \begin{bmatrix} \frac{MIDI\,time(x_{t}^{s})}{Division(x_{t}^{s})}\cdot Tempo(x_{t}^{s}) \mid t=1..T \end{bmatrix}$\Comment{match timesteps to absolute times in $x^{s}$}
    \State $abs\_time\_seg \gets [(i, i+9) \mid i=1..L-9]$\Comment{create absolute time segments}
    \State $song\_segments \gets [x^{s, k} = x^{s}[i:j] \mid (abs\_time\_x^{s}[i], abs\_time\_x^{s}[j]) \, in \, abs\_time\_seg]$\Comment{create 10 second segments of $x^{s}$}
    \State $seg\_ratings \gets [\hat{r}_{us}^{k} = SeER(x^{s, k}, U_{u}) \mid x^{s, k} \, in \, song\_segments]$\Comment{predict ratings for each segment}
    \State $x^{s, exp}_{u} \gets song\_segments[argmax_{k}({seg\_ratings})]$\Comment{determine explainability segment}
    \EndProcedure
  \end{algorithmic}
\end{algorithm}

In order to illustrate the SeER recommendation and explainability processes, we compute the top 5 recommendations for User number 1000 whose top 5 rated songs are listed in Table~\ref{tab:user4_ratings}. The top 5 recommendations with explanations for this user are shown in Table~\ref{tab:user4_recommendations}. The explanations are presented with the start and end times of the 10-second samples in $\mu s$. We provide a link to a video\footnote{\url{https://drive.google.com/file/d/1imlh4nPFhXetE1jCzPkGPm8XRxUxcRJR/view?usp=sharing}} demo where these explainability segments can be heard.

\begin{table}
\caption{Top 5 normalized ratings of User 1000 in the dataset. \label{tab:user4_ratings} }
\begin{center}
\resizebox{\linewidth}{!}{
\begin{tabular}{ l c c c c}
\hline
Artist name                       & Title                                                & Genres & Rating \\ \hline \hline
Dido                               & White Flag                                           & Pop, Hip-hop & 5      \\
Travie McCoy                       & Billionaire {[}ft. Bruno Mars{]} & Pop & 5      \\
Dido                               & Thank You                                            & Pop & 4      \\
Alicia Keys ft. Adam Levine  & Wild Horses                                          & Neo-soul & 4      \\
Michael Bubl\'e                      & Put Your Head On My Shoulder         & Easy listening & 2      \\
\hline
\end{tabular}
}
\end{center}
\end{table}

\begin{table}
\caption{Example of top 5 recommendation predicted by SeER to user 1000 (partially listed in Table~\ref{tab:user4_ratings}). The explanations are represented by the start and end times of the 10-second samples in $\mu s$. \label{tab:user4_recommendations} }
\begin{center}
\resizebox{\linewidth}{!}{
\begin{tabular}{ l c c c c c}
\hline
          &                                  &  & Predicted   &  \\
Artist name          & Title                                 & Genres & rating   & Explanation \\ \hline \hline
Andreas Johnson       & Glorious                               & Alternative/Indie, Pop & 5.360812 & (130074061.0, 139999986.0) \\
The Knack             & My Sharona                                & Rock & 5.346163 & (11172411.0, 20937925.8)   \\
Cat Stevens           & Trouble                                & Singer-songwriter & 5.330237 & (24230213.1, 33972849.8)   \\
CoCo Lee              & Before I Fall In Love                    & Contemporary R\&B & 5.314626 & (126034512.0, 135942920.0) \\
Red Hot Chili Peppers & Blood Sugar Sex Magik  & Alternative/Indie & 5.290801 & (248107860.0, 257837580.0) \\
\hline
\end{tabular}
}
\end{center}
\end{table}

\section{Offline experimental evaluation}

In this section, we describe an offline evaluation pipeline aiming to assess the recommendation performance and capabilities of our model.

\subsection{Experimental Setting}

We used the same 80/20\% train/test split for all the experiments in order to be consistent when comparing two models or when reproducing an experiment. Due to computational and time constraints, we trained all the models in 20 epochs and evaluated the results in terms of recommendation ranking using Mean Average Precision at cutoff K (MAP@K). Furthermore, in order to assess the statistical significance when comparing two models, we replicated each experiment 5 times and applied statistical tests.

\subsection{Hyperparameter Tuning}

We fixed the number of sequence model layers to 1 and the batch size to 500 because of memory constraints. Also, we relied on the Adaptive Moment Estimation (Adam) \cite{DBLP:journals/corr/KingmaB14} optimizer because it  yields a relatively fast convergence and adapts the learning rate for each parameter \cite{He:2017:NCF:3038912.3052569}. Finally, we tuned the number of latent features from 50 to 200 with increments of 50 and the sequence model type by trying RNN, GRU and Long Short-Term Memory \cite{Hochreiter:1997:LSM:1246443.1246450} (LSTM) networks. We relied on a greedy approach, that consists of varying the hyperparameters one by one independently from each other. We started by initializing the sequence model type to LSTM and tuned the number of latent features. Then, we varied the sequence model type. The results are presented in Table~\ref{tab:hyperparameter_tuning}. We obtained the best performance with 150 latent features and GRU.

\begin{table}
\caption{Hyperparameter tuning results: MAP@10 on the test data after 20 epochs. Best results (in bold) obtained, first, with 150 latent features, then, with GRU. \label{tab:hyperparameter_tuning} }
\begin{center}
{
\begin{tabular}{ l c c}
\hline
Hyperparameter & Value & MAP@10\\ \hline \hline
\# of latent features & 50 & 0.1236\\
 & 100 & 0.1424\\
 & \textbf{150} & \textbf{0.1433}\\
 & 200 & 0.1425\\ \hline
Sequence model type & LSTM & 0.1433\\
 & RNN & 0.0973\\
 & \textbf{GRU} & \textbf{0.1437}\\
\hline
\end{tabular}
}
\end{center}
\end{table}

\subsection{Research Questions}

To evaluate the prediction ability of our model, we made both wide and narrow comparisons. For the wide comparison, we matched our model to baseline recommender systems regardless of their types and data nature. On the other hand, the narrow comparison consists of comparing our model to its closest competitors which are state of the art hybrid song recommender systems. This leads us to formulate our first two research questions: \textbf{RQ1: }How does our model compare to baseline recommender systems? and \textbf{RQ2: }How does our model compare to state of the art (SOTA) hybrid song recommender systems? Also, SeER can be seen as an updated version of MF with the item embedding matrix being replaced with the output of a sequence model that takes as input our preprocessed song content data. Thus, we assess the importance of the way we use the content data by comparison to MF in the third research question: \textbf{RQ3: }What is the importance of our use of the content data? Finally, we assess whether our explanations share similar characteristics. The intuition behind this is that the shared characteristics may be interpreted as user preferences captured and incorporated in the explanations. This would indicate that the explanations represent the preferences of the user and are not an artificial product of the model. This is translated in \textbf{RQ4: } Do the personalized explanations share similar characteristics?

\subsection{RQ1: How does our model compare to baseline recommender systems?}

The baseline recommender systems we used for comparison are:

\begin{itemize}
    \item \textbf{Matrix Factorization \cite{8066567}: }One of the most used collaborative filtering techniques and basis of a large number of recommender systems including ours. We used the same number of latent factors as our model which is 150.
    \item \textbf{NeuMF \cite{He:2017:NCF:3038912.3052569}: }State of the art collaborative filtering technique that combines Generalized Matrix Factorization \cite{He:2017:NCF:3038912.3052569} (GMF) and Multi-Layer Perceptron \cite{Cun88atheoretical} (MLP). We replaced its output layer with a dot product and used MSE as a loss function because we are working with ratings. We used three hidden layers for MLP and 150 latent features for all embedding matrices.
    \item \textbf{ItemPop \cite{Rendle:2009:BBP:1795114.1795167}: }Most popular item recommendation. Used to benchmark the recommendation performance.
\end{itemize}

We present the results obtained with each model in Table~\ref{tab:rq1_results}. Our model yields an average MAP@10 of 0.1437 which is higher than all the other methods. It also has the benefit of being explainable. Furthermore, we validated our results with ANOVA \cite{fisher1925statistical} and Tukey \cite{Haynes2013} tests. All the p-values were lower than 0.01 meaning that our model performs significantly better than all the other models. 
\textbf{Note that comparing our model to MF can be considered as an ablation study that aims to prove the importance of the sequence model layers and the use of the content data. In fact, replacing the sequence model layers with an embedding layer reduces our model to MF.}

\begin{table}
\caption{Comparison of SeER with baseline models: MAP@10 results after 20 epochs for 5 replicates. \label{tab:rq1_results} }
\begin{center}
{
\begin{tabular}{ l c c c c c c c}
\hline
Replicate & SeER & MF & NeuMF & ItemPop\\ \hline \hline
1  & \textbf{0.1436} & 0.1289 & 0.1314 & 0.0778\\
2  & \textbf{0.1481} & 0.1292 & 0.1303 & 0.0778\\
3  & \textbf{0.1399} & 0.1285 & 0.1366 & 0.0778\\
4  & \textbf{0.1453} & 0.1266 & 0.1376 & 0.0778\\
5  & \textbf{0.1414} & 0.1288 & 0.1378 & 0.0778\\ \hline
Average & \textbf{0.1437} & 0.1284 & 0.1347 & 0.0778\\
\hline
\end{tabular}
}
\end{center}
\end{table}

\subsection{RQ2: How does our model compare to SOTA hybrid song recommender systems?}

The most related hybrid song recommender system we found is \cite{DBLP:journals/corr/OramasNSS17}. It applies MF-, Convolutional Neural Network \cite{Lecun98gradient-basedlearning} (CNN)- and MLP-based \cite{Cun88atheoretical} models on play counts, audio spectrograms and artist biographies to generate recommendations. The dataset used is a subset of the MSD that overlaps with ours. Their dataset includes around 1M users and 328,821 songs. We compared our model directly to the results in \cite{DBLP:journals/corr/OramasNSS17} using the same evaluation process that they used. Although comparing two models on overlapping datasets is unconventional, the results can give us an idea about the ranges in which the ranking performances of the two models are. The best performing configuration, MM-LF-LIN \cite{DBLP:journals/corr/OramasNSS17}, presents a MAP@500 of 0.0036, which is significantly lower (ANOVA p-value < 0.01) than our average performance of 0.1438 as presented in Table~\ref{tab:rq2_results}.

\begin{table}
\caption{Comparison of SeER with MM-LF-LIN \cite{DBLP:journals/corr/OramasNSS17} on an overlapping dataset. Our model's performance is assessed with MAP@500 after 20 epochs with 5 replicates. \label{tab:rq2_results} }
\begin{center}
{
\begin{tabular}{ l c c}
\hline
Replicate & SeER & MM-LF-LIN\\ \hline \hline
1  & 0.1438 & 0.0036\\
2  & 0.1483 & -\\
3  & 0.1400 & -\\
4  & 0.1455 & -\\
5  & 0.1415 & -\\ \hline
Average & \textbf{0.1438} & 0.0036\\
\hline
\end{tabular}
}
\end{center}
\end{table}

\subsection{RQ3: What is the importance of our use of the sequential content data?}

We can assess the importance of using the sequential song content data as follows:
\begin{itemize}
    \item First, as we already proved in RQ1, the content data helped improve the recommendation performance since our model performs significantly better than pure rating-based MF.
    \item Second, the content data allowed us to solve the item cold start problem because the item data comes from both the song's MIDI content data and the user ratings. Thus, items with no ratings can be recommended by relying solely on their content.
    \item Finally, the sequential nature of our content data, in addition to the structure of our model, allowed us to generate 10-second instrumental explanations making recommendation more transparent. The explanations are evaluated in RQ4 below.
\end{itemize}

\subsection{RQ4: Do the personalized explanations share similar characteristics that capture user-preferences?}

In order to validate the 10-second segment explanations offline, we tried to determine, for every user, whether their personalized explanations share common characteristics. Explanations that share common properties are likely to be generated based on captured hidden preferences of the user. Hence, these explanations may represent the most important sections of the recommended songs. In that case, the explanations may not be just artefacts. To study the latter property, we propose two approaches based on analysis of the song content similarities and tags respectively.

\subsubsection{Content-based validation}

Given that we have the content of the explanations, we first relied on similarity measures to prove that they share similar characteristics. We randomly selected 100 users as our test sample. For every test user, we use our model to generate the top 5 recommendations with explanations and compute the average Dynamic Time Warping (DTW) \cite{salvador2007toward} distance between the explanations (DTWe). DTW is a powerful distance measure between multidimensional time series that do not necessarily have the same size. To generate the average DTW distance between two lists of multidimensional time series, we compute the DTW distance matrix between them and take the average of all the values in the matrix. In the case of DTW distances between explanations, both lists are similar and include the song arrays of the generated 5 explanations. To compare, we selected a random 10-second segment from every recommended song and computed the average DTW distance between these 5 segments (DTWr) for every user. We compare to average DTW distances between 10-second segments instead of between the whole recommended songs to avoid any bias coming from the different song lengths. Finally, we considered the problem as a Randomized Complete Block Design (RCBD) \cite{olsson1978randomized} and applied a Tukey test \cite{Haynes2013} for pairwise comparison. The null hypothesis is whether the averages over all the users of the average DTW distances between the explanations and between the random segments are similar. For simplicity, we will call these two entities "Avg. DTW between explanations" (or DTWe) and "Avg. DTW between random segments" (or DTWr). We show these average values with the 95\% Confidence Intervals (CIs) of the difference (DTWe - DTWr) and the statistical test results in Table~\ref{tab:content_based_validation_results}.
We notice that Avg. DTW between explanations is significantly smaller than Avg DTW between random segments (p-value<0.05 and 0 is not in the Confidence Interval). This means that for each user, we can assert with 95\% confidence that the explanations are significantly close to each other compared to the random segments from the recommendations. \textbf{Thus, we can assert that our generated 10-second segment explanations share common characteristics which are likely to represent the preferences of the user.}

\begin{table}
\caption{Significance testing with 95\% confidence of the difference between Avg. DTW between explanation and Avg. DTW between random segments: The explanations are significantly close to each other compared to the random segments. This means that the explanations capture and share common characteristics that are likely to represent the user's preferences. \label{tab:content_based_validation_results} }
\begin{center}
\resizebox{\linewidth}{!}
{
\begin{tabular}{ l c c c c c c c}
\hline
Avg. DTW between & Avg. DTW between & 95\% CI of the difference & Adjusted\\
explanations (DTWe) & random segments (DTWr) & (DTWe - DTWr) & p-value\\ \hline \hline
\textbf{22,844.1} & 24,820.9 & (182, 3,771) & \textbf{0.031}\\
\hline
\end{tabular}
}
\end{center}
\end{table}

\subsubsection{Tag-based validation}

Tags can capture an item's properties. In the case of songs, they can include genres, the era, the name of the artist or subjective emotional descriptions. We used the tags from the "Last.fm" dataset \cite{Bertin-Mahieux2011} provided with the MSD. These tags are available for almost every song in the MSD and amount to 522,366 tags \cite{Bertin-Mahieux2011}. In our dataset, we selected the songs that intersect with the "Last.fm" dataset and filtered the tags that occur in at least 100 songs in order to remove noisy tags. We obtained 4,659 songs with 227 tags. From the users that interacted with these songs, we filtered the ones that have at least 10 liked songs. In fact, we made the assumption that a rating strictly higher than 3 means that the user likes the song. Next, we randomly selected 100 users as our test sample. For every user, we determined the Top 1, 2 and 3 preferred tags, based on the tags of their liked songs, and generated the top 5 recommendations with explanations. Our objective is to determine how much the personalized recommendations and explanations match the preferred tags of every user. Thus, we need to determine the tags of both the recommendations and the explanations, which are not necessarily in the tags dataset. To cope with this issue, we trained a multi-label classification model on our tags dataset to predict the tags of the recommendations and explanations. The classification model is basically a sequence model layer with 20\% dropout followed by Multilayer Perceptron (MLP) \cite{Popescu:2009:MPN:1639537.1639542} layers with ReLU activation functions and an output layer with 227 nodes, corresponding to the 227 classes, each with a Sigmoid activation function. The model is trained to optimize the Binary Cross-entropy loss to predict the probability of each tag individually in every node \cite{lapin2017analysis}. To tune our model's hyperparameters, we started with an LSTM layer followed by the output layer. We tuned the size of the hidden state from 100 to 500 with an increment of 100. Then, we tuned the number of MLP hidden layers from 1 to 5. We chose the number of nodes in the hidden layers to be the optimal size of the hidden state, which is 300. Finally, we tuned the sequence model type of the first layer by additionally trying RNN and GRU. The best model has one LSTM layer with a hidden state size of 300 followed by 4 MLP layers of the same size and, finally, the output layer. We reached a performance of 93.4\% accuracy and respectively 51.8\%, 61.9\% and 67.7\% top-1, top-2 and top-3 categorical accuracy with 5-fold cross validation. We used top-k categorical accuracy \cite{lapin2017analysis} because we are interested in correctly predicting the existing tags in a sparse target. We used our trained classifier to predict the tags of all the recommendations and explanations for all the users. Then, we calculated the Average Percentage Matching of the recommendations and explanations with the top 1, 2 and 3 user preferred tags. We define the Percentage Matching of a list of songs $S$ with the top $k$ preferred tags $T_{k}(u)$ of a user $u \epsilon U$ as the percentage of songs from $S$ including at least one of the top $k$ preferred tags $T_{k}(u)$, as follows:

\begin{equation}
    \% \, Matching(S, T_{k}(u)) = \frac{100}{|S|} |\{s \epsilon S | Tags(s) \cap T_{k}(u) \neq \emptyset\}|
\end{equation}

$Tags(s)$ is the set of tags of the song $s$. In our case, the set of tags of a recommendation or an explanation is predicted using the multi-label classification model. The Average Percentage matching is the average of the Percentage Matching over all the test users:

\begin{equation}
    Avg \, \% \, Matching(S, U, k) = \frac{100}{|U|} \sum_{u=1}^{|U|} \% \, Matching(S(u), T_{k}(u))
\end{equation}

$S(u)$ in our case is either the set of recommendations or explanations of user $u$. We varied $k$ and considered every problem as a RCBD \cite{olsson1978randomized}. We applied Tukey tests \cite{Haynes2013} for pairwise comparison. The null hypothesis for every test is whether the average percentage matchings of the recommendations and of the explanations with the top $k$ liked songs (Avg\%Matching(rec., U, k) and Avg\%Matching(exp., U, k) respectively) are equal. We show the two average percentage matching values with the corresponding 90\% CIs of the differences (Avg\%Matching(rec., U, k) - Avg\%Matching(exp., U, k)) and adjusted p-values of the Tukey tests in Table~\ref{tab:percentage_matching_results}. We notice that for all k, the explanations match the preferred tags of the users more than the recommendations. The difference is significant for k = 1 (CI of the difference does not include 0 and p-value<0.1). However, starting from k = 2, the difference becomes insignificant as both the recommendations and explanations start matching the top k preferred tags comparably well, but still with a slight advantage for the explanations. \textbf{This means that the explanations share similar properties which are more in accordance with the preferred tags of the users than even the overall recommendations.} Assuming that the tags represent the genres, if the user's preferred genre is, for instance, ``Rock" and a ``Pop" song gets recommended, the explanation of that song is likely to be a ``Rock" segment of the song. We show an illustrative example of a user from our test sample in Table~\ref{tab:example_explainability_validation}.

\begin{table}
\caption{Significance testing with 90\% confidence of the difference between the Avg \% Matching of recommendations and explanations with user top k preferred tags. \label{tab:percentage_matching_results} }
\begin{center}
\resizebox{\linewidth}{!}
{
\begin{tabular}{ l c c c c c c c}
\hline
 & Avg\%Matching & Avg\%Matching & 90\% CI & Adjusted \\
k & (rec., U, k) & (exp., U, k) & of the difference & p-value\\ \hline \hline
1  & 84.24\% & \textbf{84.85}\% & (-0.01181, -0.00031) & \textbf{0.083}\\
2  & 90.71\% & \textbf{90.91}\% & (-0.00537, 0.00133) & 0.320\\
3  & 94.75\% & \textbf{94.95}\% & (-0.00537, 0.00133) & 0.320\\
\hline
\end{tabular}
}
\end{center}
\end{table}

\begin{table}
\caption{Example of a test user (\#26647) where the explanations match the favorite tags more than the recommendations: The first recommended song is a "pop" song. However, the explainability segment is both "pop" and "rock" which matches the favorite tags of the user better than the recommendation itself. \label{tab:example_explainability_validation} }
\begin{center}
\resizebox{\linewidth}{!}
{
\begin{tabular}{ l c c c c c c c}
\hline
Recommendation & \multicolumn{2}{c}{Recommendation tags} & \multicolumn{2}{c}{Explanation tags} \\ \hline \hline
1 & \multicolumn{2}{c}{\textbf{pop}} & \multicolumn{2}{c}{pop, rock}\\
2  & \multicolumn{2}{c}{pop, rock} & \multicolumn{2}{c}{pop, rock}\\
3  & \multicolumn{2}{c}{pop, rock} & \multicolumn{2}{c}{pop, rock}\\
4  & \multicolumn{2}{c}{pop, rock} & \multicolumn{2}{c}{pop, rock}\\
5  & \multicolumn{2}{c}{pop, rock} & \multicolumn{2}{c}{pop, rock}\\

User top 3 tags (sorted) & \multicolumn{4}{c}{\textbf{rock}, pop, favorites}\\ \hline
k & & 1 & 2 & 3\\ \hline \hline
$\% \, Matching(rec., T_{k}(u))$ & & \textbf{80\%} & 100\% & 100\%\\
$\% \, Matching(exp., T_{k}(u))$ & & 100\% & 100\% & 100\%\\ \hline
\end{tabular}
}
\end{center}
\end{table}

\section{Conclusion}

We proposed a hybrid song recommender system that uses both ratings and song content to generate personalized recommendations accompanied with short MIDI segments as explanations. We made recommendation more transparent while relying on powerful deep learrning models. Our experiments demonstrated that our architecture performs significantly better than both baseline and SOTA hybrid song recommender systems. Moreover, we validated the effectiveness of the way we integrate the content data and solved the item cold start problem which is a notorious limitation of Collaborative Filtering techniques. Finally, we validated our explainability approach by showing that the personalized explanations are able to capture properties that are in accordance with the preferences of the user. Our approach has limitations such as the slow training time and the user cold start problem. In the future, we plan to extend our methods to more complex and challenging modalities such as images and videos.

%
\bibliographystyle{ACM-Reference-Format}
\bibliography{sample-base}

\end{document}